\documentclass{aa}
 
\usepackage[dvips]{color}
\usepackage{graphicx}
\usepackage{txfonts}
\usepackage{rotating}
\newcommand{\kms}{$\rm{km}\,\rm{s}^{\rm -1}$}
\newcommand{\masyr}{${\rm mas}\,{\rm yr}^{-1}$}

\begin{document}

\title{Absolute proper motion of the Galactic open cluster
M~67\thanks{Based on data acquired with the Large Binocular
Telescope (LBT) at Mt. Graham, Arizona, under the Commissioning of the
Large Binocular Blue Camera.  The LBT is an international
collaboration among institutions in the United States, Italy and
Germany. LBT Corporation partners are the University of Arizona on
behalf of the Arizona university system; Istituto Nazionale di
Astrofisica, Italy; LBT Beteiligungsgesellschaft, Germany,
representing the Max-Planck Society, the Astrophysical Institute
Potsdam, and Heidelberg University; the Ohio State University, and the
Research Corporation, on behalf of the University of Notre Dame,
University of Minnesota and University of Virginia; and on
observations obtained at the Canada-France-Hawaii Telescope (CFHT),
which is operated by the National Research Council of Canada, the
Institut National des Sciences de l'Univers of the Centre National de
la Recherche Scientifique of France, and the University of
Hawaii.}}

\author{Bellini, A.\inst{1,}\inst{2,}
\thanks{Visiting PhD Student at
STScI under the \textit{``2008 graduate research assistantship''}
program.}
\and
Bedin, L. R.\inst{2}
\and
Pichardo, B.\inst{3}
\and
Moreno, E.\inst{3}
\and
Allen, C.\inst{3}
\and
Piotto, G.\inst{1}
\and 
Anderson, J.\inst{2}\\ 
}
%
%\offprints{Bellini, A.}
%
\institute{Dipartimento di Astronomia, Universit\`a di Padova, Vicolo
dell'Osservatorio 3, I-35122 Padova, Italy 
%\\
%\email{[andrea.bellini;giampaolo.piotto]@unipd.it}
%
\and Space Telescope Science Institute, 3700 San Martin Drive,
Baltimore, MD 21218, USA 
%\\ 
%\email{[bellini;bedin;jayander]@stsci.edu}
%
\and Instituto de Astronom\'ia, Universidad Nacional Aut\'onoma de
M\'exico, Apartado Postal 70-264, 04510-M\'exico DF, Mexico
%\\
%\email{[barbara;edmundo;chris]@astroscu.unam.mx}
}
\date{Received 15 Dicember 2009 / Accepted 23 February 2010}
%________________________________________________________________

\abstract{
We derived the absolute proper motion (PM) of the old,
solar-metallicity Galactic open cluster M~67 using observations
collected with CFHT (1997) and with LBT (2007).  About 50 galaxies
with relatively sharp nuclei allow us to determine the absolute PM of
the cluster.  We find $(\mu_\alpha\cos\delta,\,\mu_\delta)_{\rm
J2000.0}$ $=$ $(-9.6\pm1.1,\,-3.7\pm0.8)$ \masyr.  By adopting a
line-of-sight velocity of $33.78 \pm 0.18$ \kms, and assuming a
distance of 815$\pm$50 pc, we explore the influence of the Galactic
potential, with and without the bar and/or spiral arms, on the
galactic orbit of the cluster.
}


\maketitle

%%%%%%%%%%%%%%%%%%%%%%%%%%%%%%%%%%%%%%%%%%%%%%%%%%%%%%%%%%%%%%%%%%%%%%

\section{Introduction}
\label{sec:1}

The solar-metallicity Galactic open cluster M~67 (NGC~2682) is among
the most-studied Galactic open clusters. Still, its absolute proper
motion (PM) remains poorly constrained.

We applied for the first time on ground-based multi-epoch CCD
wide-field images the PM techniques developed in Anderson et al.\
(2006, Paper~I), Yadav et al.\ (2008, Paper~II), Bellini et al.\
(2009, Paper~III) to define an absolute reference frame using
background faint galaxies.  In the next section we describe the data
set and the measurements, and a final section is dedicated to the
study of the orbit of M 67 within the Galaxy under different
assumptions for the Galactic potential.

\section{Observations, data reduction, proper motions}
\label{sec:2}

Two data sets -- collected with two different telescopes and at two
different epochs -- were used to measure the PM of objects in the field
of M~67.

As the first epoch (hereafter, epoch\ 1) we employed images taken on
Jan.\ 10--13, 1997, at the CFHT 3.6m telescope.  These images were
first presented by Richer et al.\ (\cite{richer98}). We took only a
subsample of this data set, specifically 15 exposures of 1200 s in the
$V$ filter, with a median value for seeing and airmass of 1$\arcsec$
and 1.2 respectively.  Each image was collected with the UH8K camera
(8 CCDs, 2K$\times$4K pixels each, with an average scale of 210 mas\
pixel$^{-1}$) covering a field-of-view (FoV) of
$\sim$29$^\prime$$\times$29$^\prime$.

The second-epoch (epoch\ 2) data were collected between Feb.\ 16 and
Mar.\ 18 2007 and consist of\ 56 images of 180 s exposures in the
$B$-band filter and 42 exposures of 100--110 s in the $V$-band,
obtained with the LBC-blue camera (4 CCDs, 2K$\times$4.5K pixels each,
FoV of $\sim$24$^\prime$$\times$26$^\prime$).  A large dither pattern
($\sim$30\% of the FoV) was employed for both filters.  Median seeing
and airmass were 1$\arcsec$ and 1.1 for the $V$, and $1\farcs3$ and
1.1 for the $B$ images.  We selected only $B$ images for finding
objects, while those in $V$ were taken to provide all the astrometric
information.  A more extensive description of the LBT data set is
given in three companion papers (Bellini \& Bedin \cite{bb10}, Bellini
et al.\ {\cite{bellini10}, and Bellini et al.\ in prep.).  Procedures
and algorithms used to derive the list of objects, star positions and
fluxes, and PMs are those explained in great detail in Paper I and in
Anderson et al.\ (\cite{anderson08}, A08).  Below we briefly describe
these reduction procedures, which were organized in three steps.

\begin{figure}[th!]
\centering
\includegraphics[width=\columnwidth]{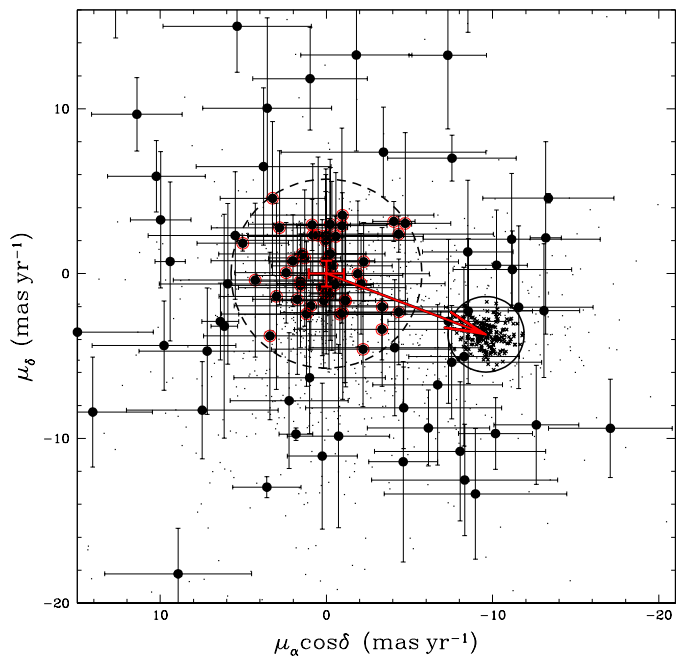}
\caption{Vector-point diagram of the absolute PMs, in equatorial
coordinates (J2000.0).  Small crosses are M 67 members (within the
solid circle), filled circles are the reference galaxies, 47 of which
(marked with red open circles) were taken to compute the centroid of
the galaxy distribution.}
\label{fig:1}
\end{figure}

In the first step, we employed the software described in Paper~I to
obtain PSFs, star positions and fluxes in each chip of each exposure
and for the best sources (bright, isolated, with a stellar profile).
We corrected LBC raw positions for geometric distortion (GD), as
described in Bellini \& Bedin (\cite{bb10}), and we used these
corrected positions to register all of the LBC single-chip images into
a common distortion-free reference frame (the master frame) by means
of linear transformations.  With this master frame we derived the GD
correction also for the UH8K positions, using the same technique (used
also in Bellini \& Bedin \cite{bb09}).

The second step consisted of producing the list of objects (the target
list).  We took every single local maximum detected within each
individual chip of each LBC $B$ image to build a peak-map image (A08).
The peak map consists of an image with the same pixel scale of the
master frame, where we added 1 to a pixel count each time a local
maximum, measured in a given image, fell within that pixel (once
transformed with the aforementioned linear transformations).  A
3$\times$3 box-car filter is applied to the peak map, and in our final
list we considered as significant any object 3.2$\sigma$ (where
$\sigma$\ is the rms of the peak-map background) above the local
surrounding.  Our list contains objects generating a local maximum in
at least 40\% of the images covering the patch of sky where the
maximum is found.  We additionally required a minimum coverage of 10
$B$ images.

With the $B$ images we generated the target list because they are more
numerous and have a lower background compared to $V$ images.  In
addition, large dithers helped us to produce a list of solid
detections even for the faintest sources.  We further purged our list
from excessively faint sources, keeping only objects with at least 100
DN above the local background and which showed up in at least one
observation for each of the two $V$-filter epochs.  We also excluded
objects within chip \# 2 and 4 of the UH8K camera, because these chips
are highly affected by charge transfer inefficiency.  As a
consequence, the useful FoV is reduced by 25\%.

In the last step we derived PMs (see Sect.~7 of Paper~I for more
details):\ We measured each object in the target list in each chip of
each $V$ exposure where it could be found, using PSF-fitting to get a
chip-based flux and a GD-corrected position.  Then we organized the
images in pairs of one image from the first and one from the second
epoch.  For each object in each pair we computed the displacement (in
the reference frame) between where epoch\ 1 predicts the object
position in epoch\ 2, and the actually-observed position in epoch\ 2.
Multiple measurements of displacements for the same object were then
used to compute average displacements and rms.

It is clear that to make these predictions, we needed a set of objects
as a reference to compute positional transformations between the two
epochs for each source.  The cluster members of M~67 were a natural
choice, as their internal motion is within our measurement errors
($\sim$0.2 mas yr$^{-1}$ Girard et al.\ \cite{girard89}), providing an
almost rigid reference system with the common systemic motion of the
cluster.

We initially identified cluster members according to their location in
the $V$ vs.\ $B-V$ color-magnitude diagram.  We took only main
sequence (MS) stars to transform each exposure into the master frame,
because it spans a narrow range in color\footnote{Note that it was not
possible to use the high-probability members derived in Paper~II as
starting reference members, because there are not enough stars in
common between the two catalogs. Indeed, the PM catalog of Paper~II
includes stars in the range magnitude 9$<$$V$$<$21 (with reliable PM
measurements down to $V$$\sim$19), while our new PM determinations are
in the magnitude range 18$<$$V$$<$26.}.  By predominantly using
cluster members, we ensured the PMs to be measured relative to the
bulk motion of the cluster.  We iteratively removed from the member
list those objects with a field-type motion (i.e., with a PM larger
than 2.3 mas\ yr$^{-1}$ from the mean M~67 motion), even though their
colors may have placed them near the cluster MS.  This particular cut
value in the PMs represents a compromise between loosing (poorly
measured) M~67 members and including field-type objects.  Our final
member list contains 209 color- and PM-selected objects.  In order to
minimize the influence of any uncorrected GD residual, PMs for each
object were computed with a local sample of members; specifically the
25 (at least) closest ($r$$<$3$^\prime$), well-measured cluster stars
(see Paper~I for more details).

\begin{figure}[t!]
\centering
\includegraphics[width=\columnwidth]{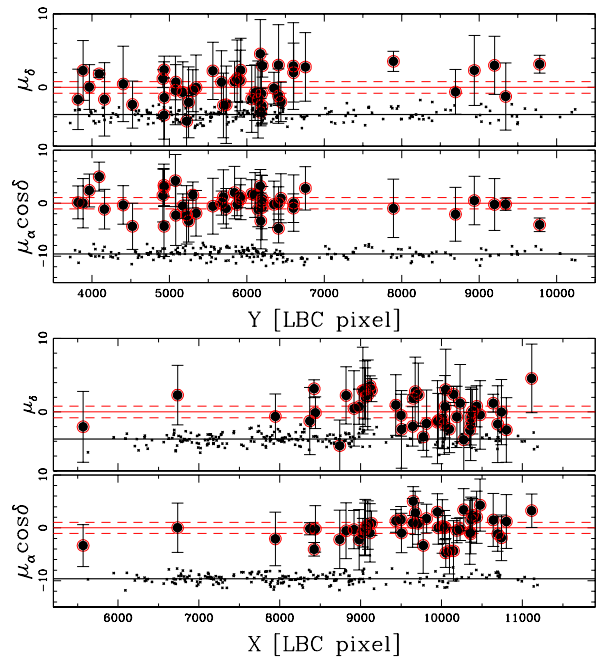}
\caption{From top to bottom, $\mu_\delta$ and $\mu_\alpha \cos\delta$
as a function of X and Y. When PMs are plotted as a function of their
position in the master frame (X,Y), no clear systematic errors appear.
Note that two chips of UH8K are not used (Y$>$7000, for X$<$5500 \&
X$>$9500), and that the cluster center is at location
(X,Y)$\simeq$(7500,7000).  These two effects both create gaps in the
galaxy spatial distribution.}
\label{fig:2}
\end{figure}

\begin{figure}[t!]
\centering
\includegraphics[width=\columnwidth]{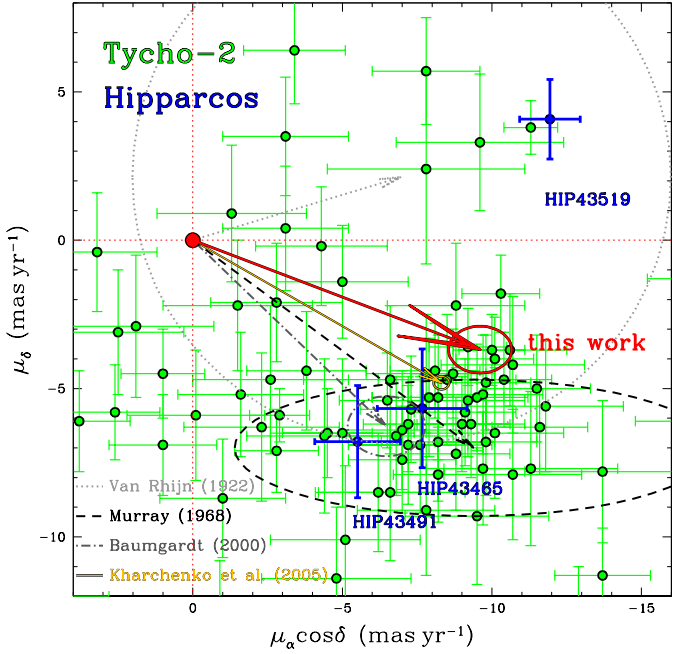}
\caption{Comparison of our derived absolute PMs with values from the
literature.  The three blue points are Hipparcos PM determinations,
while green points are from the Tycho-2 catalog.  Our M~67 absolute PM
is indicated with a red arrow, and the ellipse in its point shows our
uncertainty.  The other four arrows in different colors and types
refer to previous values.}
\label{fig:3}
\end{figure}

Then we matched our master frame with the sources in the Digital Sky
Survey (DSS) to compute scale and orientation for our master frame.
We needed to know scale and orientation with an error of $\sim$
1\%. Even if only saturated stars could be matched with the DSS
catalog, the used sample was good enough for this purpose.  We divided
our displacements for the time baseline between the two epochs (10.13
years) to obtain PMs in units of mas yr$^{-1}$.  We kept in our final
PM list only objects with at least two measurements of the
displacement, and with PM rms $<$7 mas\ yr$^{-1}$ in each coordinate.
Finally, we corrected our PMs for differential chromatic refraction
(DCR) effects, as done in Papers~I \& III, using M~67 white dwarfs and
MS stars.  We note that DCR corrections were always below 0.2 mas\
yr$^{-1}$\ $(B-V)$$^{-1}$.

On the basis of the PM dispersion of members ($\sim$0.9 mas\
yr$^{-1}$) with respect to their mean (which reflects our measurement
errors) we estimated the uncertainty on the adopted member reference
system to be $\pm$(0.1,0.1) mas\ yr$^{-1}$.  (Note that our estimate
of the member dispersion is somehow biased because of the 2.3 mas\
yr$^{-1}$ membership selection criterion we adopted in our
proper-motion determination. Hence, the dispersion could be
underestimated. Nevertheless, it still provides a good indication of
the involved errors.)  We were still left with the problem of finding
the zero point of our PMs -- which for now were only relative to the
cluster's mean motion and not to an absolute reference system.
Unfortunately, it was not possible to directly link our proper motions
to the UCAC3}\footnote{{\sf
http://www.usno.navy.mil/USNO/astrometry/optical-IR-prod/ucac/}.}
catalog (Zacharias et al.\ \cite{zacharias09}), which includes stars
mainly in the 8 to 16 magnitude range in a single bandpass between $V$
and $R$, with no overlap with our catalog.

Background Galaxies can be considered as fixed points in the sky, and
provide an excellent, and {\it directly-observable} absolute reference
system.  A visual inspection of the images revealed many such
galaxies. About 100 of them show point-like nuclei, which could be fitted
with our PSFs to measure positions and PMs (as done in Bedin et al.\
\cite{bedin03}, \cite{bedin06}).

As expected, the errors of galaxy positions are several times larger
than the typical error of star positions, and depend strongly on
galaxy morphology.  The comparison of the galaxy PMs (which should all
be the same), suggests that we are underestimating their errors (which
is not the case, for the point sources), as a result of a complex
combination of seeing and galaxy shape.  For this reason, we did not
use their weighted mean to estimate the centroid of the galaxy PMs,
but adopted an iterative $\sigma_{\mu}$-clipped average instead.

\begin{table*}[t!]
\begin{center}
\caption{Non-axisymmetric Galactic model (Pichardo et al.\
  \cite{PMME03}, \cite{PMM04}).}
\label{tbl-1}
\begin{tabular}{lcl}
\hline\hline
&&\\
Parameter & Value & References\\
&&\\
\hline
&&\\
Bar half-length                 &3.1--3.5 kpc & Gerhard (\cite{G02}) \\
Bar axial ratios                &10:3.8:2.6& Freudenreich (\cite{F98}) \\
Bar scale lengths               &1.7, 0.64, 0.44 kpc & Freudenreich (\cite{F98}) \\
Bar angle (respect to the Sun)  & $20^{\rm o}$   & Gerhard (\cite{G02}) \\
Bar mass                                  & $10^{10}\ {\rm M}_\odot$&Debattista et al. (\cite{DGS02})\\
Bar pattern speed ($\Omega_{B}$) & 30--60 km$\,$s$^{-1}$$\,$kpc$^{-1}$& ($^\ast$) \\
Spiral Arms locus                         & Bisymmetric (Logthm) & Churchwell et al. (\cite{CBM09})\\
Spiral Arms pitch angle                   & $15.5^{\rm o}$ & Drimmel (\cite{D00}) \\
Spiral Arms external limit                & 12 kpc & Drimmel (\cite{D00}) \\
Spiral Arms: exp. with scale-length     & 2.5 kpc & Disk based\\
Spiral Arms force contrast                & $\sim$ 10\% & Patsis et al. (\cite{PCG91}) \\
Spiral Arms pattern speed ($\Omega_{S}$) & 20 km s$^{-1}$ kpc$^{-1}$ & Martos et al. (\cite{MHY04})\\
&&\\
\hline
\multicolumn{3}{l}{\small{Note: ($^\ast$)~Weiner
  \& Sellwood (\cite{WS99}), Fux (\cite{F99}), Ibata \& Gilmore
  (\cite{IG95}), Englmaier \& }}\\
\multicolumn{3}{l}{\small{Gerhard (\cite{EG99}).}}\\
\end{tabular}
\end{center}
\end{table*}

\begin{figure*}[t!]
\centering
\includegraphics[width=16cm]{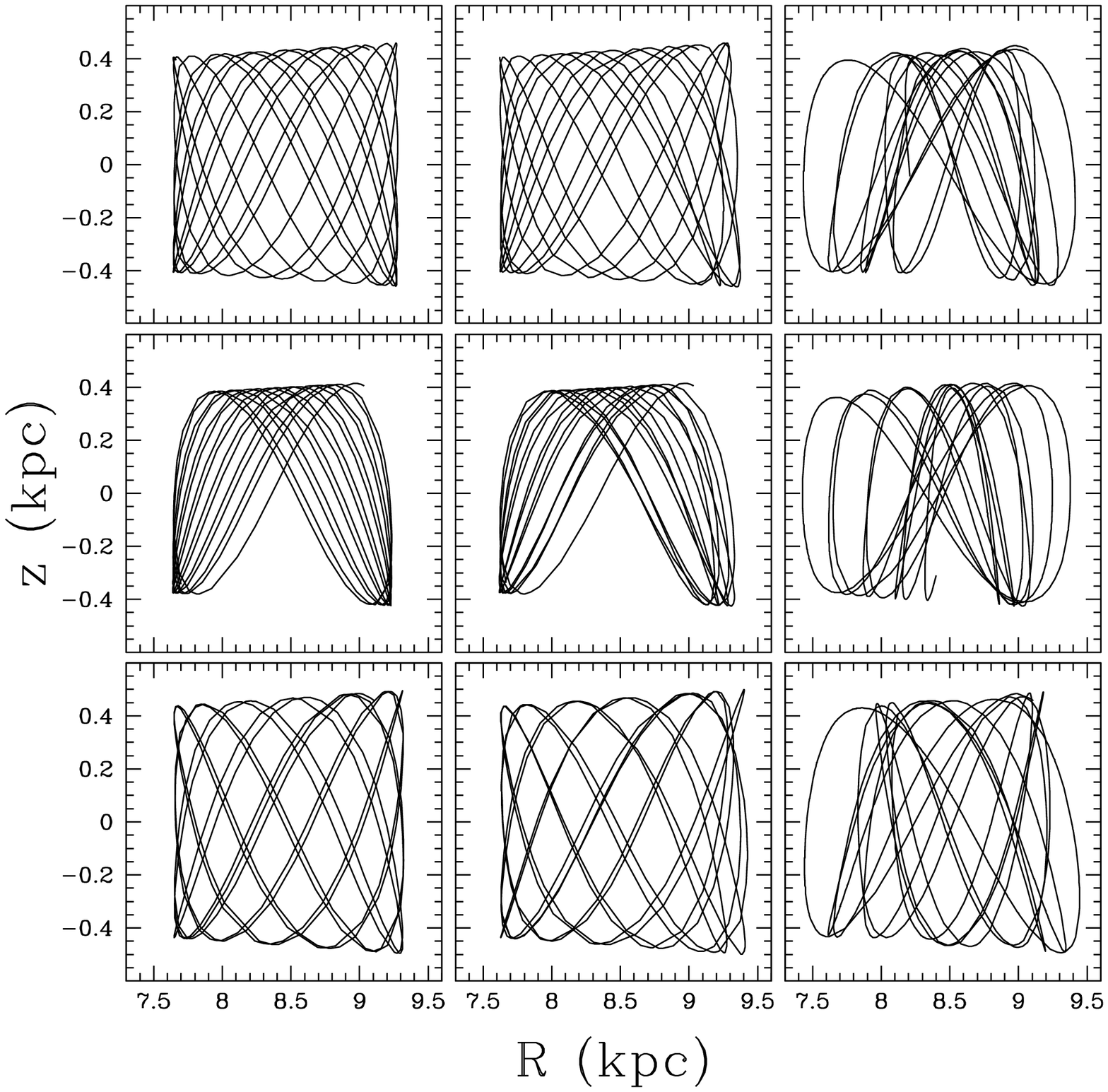}
\caption{Meridional orbits using the three heliocentric distances;
$d$=0.815 kpc (upper panels), 0.765 kpc (\textit{middle panels}), and 0.865 kpc
(\textit{lower panels}), computed in the axisymmetric (\textit{left panels}), barred
(\textit{central panels}), and with bar + spiral arms (\textit{right panels}) scaled
Galactic potential.}
\label{fig1}
\end{figure*}

We started with all the $N^{\rm G}$$\simeq$100 galaxies in the sample,
for which the motion is within 20 mas\ yr$^{-1}$ from the first-guess
mean PM of galaxies, and we iteratively estimated their PM dispersion
$\sigma_{\mu}^{\rm G}$ as the 1-D 68.27$^{\rm th}$-percentile of their
distribution around the median motion.  We then excluded from the
sample those galaxies with PM$>$2.5$\sigma_{\mu}^{\rm G}$ from the
median galaxies' PM, and we re-derived new values of
$\sigma_{\mu}^{\rm G}$, median motion and $N^{\rm G}$. We iterated
this procedure 10 times, noting that the values of median,
$\sigma_{\mu}^{\rm G}$, and $N^{\rm G}$ converged after eight
iterations.  Final values for $\sigma_{\mu}^{\rm G}$ and $N^{\rm G}$
are 2.3 mas\ yr$^{-1}$ and 47, respectively.

By adopting as the origin of the absolute PM system the final median
value of the PM of galaxies, we found for M~67 an absolute PM
(J2000.0) of
$$
(\mu_{\alpha}\cos
\delta,\mu_{\delta})=(-9.6,-3.7)\pm(1.1,0.8){\rm~mas}\,{\rm yr}^{-1},
$$
where the uncertainties come from adding in quadrature the
uncertainties of the centroid of M 67 members to the error of the
location of the centroid of the galaxies (for each coordinate
independently).

The entire sample of $\sim$100 galaxies was used to estimate the
uncertainty associated with the displacement of a single, typical
galaxy. This was calculated as the 1-D 68.27$^{\rm th}$-percentile of
the distribution of the galaxies around their median displacement (as
computed above).  Specifically: $(\sigma^{\rm
G})_{\mu_{\alpha}\cos\delta}$=7.1 mas\ yr$^{-1}$ and $(\sigma^{\rm
G})_{\mu_{\delta}}$=5.3 mas\ yr$^{-1}$.  Because we only used the most
consistent subset of 47 galaxies to derive their relative median
motion, we associated to the median of their displacements an
uncertainty that statistically takes into account only those $N^{\rm
G}$$=$47 galaxies.  This is done by reducing the error on the single
average galaxy by the factor $1/\sqrt{N^{\rm G}-1}$.  This
more-conservative approach might result in a overestimate of our
internal errors. However, considering all the uncertainties involved
in the assignment of an error to a galaxy displacement and their
limited number, this is a preferable approach.

Figure~\ref{fig:1} shows our absolute vector-point diagram (VPD) for
all the objects in the final list (small dots).  M 67 members are
marked with small crosses.  These are the stars within 2.3 mas\
yr$^{-1}$ from the MS stars' mean PM (solid circle).  Filled circles
are visually-confirmed galaxies, and are shown with our estimated
error bars.  The best galaxies (selected with the aforementioned
iterative procedure) are highlighted with red open circles.  The
Figure also shows the 2.5$\sigma^{\rm G}$=5.7 mas\ yr$^{-1}$ selection
radius (dashed circle).  The red arrow indicates our estimate of the
absolute PM of M~67.  The error bars at the base of the arrow indicate
the total uncertainty (dominated largely by the estimates in the
centroid of the relative PM distribution of the background galaxies).
In Fig.\ \ref{fig:2}, we used the same symbols as in Fig.\ \ref{fig:1}
to show that there are no clear systematic errors in the galaxy PMs as
a function of the coordinates of the master frame (X, and Y, parallel
to $\alpha$ and $\delta$, respectively).

The absolute PM of M~67 has been measured by several authors.  The
first of these determinations comes from van Rhijn (\cite{vr22}), with
$(\mu_{\alpha}\cos \delta,\mu_{\delta})=(-7,+2)\pm(9,9){\rm~mas}\,{\rm
yr}^{-1}$.  Many years later, Murray (\cite{murray68}) computed
$(\mu_{\alpha}\cos
\delta,\mu_{\delta})=(-9.4,-7.0)\pm(8.0,2.3){\rm~mas}\,{\rm yr}^{-1}$.
The more contemporary work of Baumgardt et al.\ (\cite{bau00}), which
made use of the two faint Hipparcos stars HIP-43491, HIP-43465
($V$$\sim$10), reports: $(\mu_{\alpha}\cos
\delta,\mu_{\delta})=(-6.47,-6.27)\pm(1.29,1.01){\rm~mas}\,{\rm
yr}^{-1}$.  Finally, Kharchenko et al.\ (\cite{kharchenko05}) selected
27 M~67 members on the basis of their absolute PMs, as derived from
the ASCC-2.5 catalog (Kharchenko \cite{kharchenko01}) ---which is
based on Hipparcos-Tycho family catalogs--- and on their location on
the $V$ vs.\ $B-V$ color-magnitude diagram. The obtained mean-absolute
motion of these 27 members is $(\mu_{\alpha}\cos
\delta,\mu_{\delta})=(-8.31,-4.81)\pm(0.26,0.22){\rm~mas}\,{\rm
yr}^{-1}$.

Figure~\ref{fig:3} shows (in blue) three stars in the field with
Hipparcos PMs (HIP-43491, HIP-43465 and HIP-43519, with error bars),
and the stars from the Tycho-2 catalog (in green).  In the same plot,
we mark with a red arrow our derived M~67 absolute PM, where the
estimated PM error is indicated with an ellipse. Note that both the
Hipparcos and the Tycho-2 stars are far too bright to be measured in our
survey.  Our M~67 absolute PM determination is marginally consistent
with the bulk of Tycho-2 measurements for the objects in the same
field.  This is in line with the expected accuracies for Tycho-2.  The
same figure also shows (dashed arrows) previous determinations of the
M~67 absolute motion.

We emphasize that the absolute PM presented here is based on {\it
direct} observations of background galaxies, used to define the
absolute reference frame.  It is a purely differential measurement,
which does not rely, as do previous measurements, on a complex
registration to the International Celestial Reference System through a
global network of objects.  We end this section noticing the good
agreement between the absolute PM value of M~67 as derived with the
bright sources of the Hipparcos catalog (Kharchenko et al.\
\cite{kharchenko05}) and that based on the faint, ``fix'' galaxies
(this work).

\section{The Galactic orbit of M~67}\label{orbit}

With the absolute PM of M~67 given in the previous section, its
line-of-sight velocity of 33.78$\pm$0.18 km$\,$s$^{-1}$ and its
heliocentric distance of 815$\pm$50 pc (both from Paper II), we have
computed the Galactic orbit of M~67.  We employed a Galactic potential
that includes the axisymmetric model of Allen \& Santill\'an
(\cite{AS91}) and the bar and spiral arms models of Pichardo et al.\
(\cite{PMME03}, \cite{PMM04}).  The axisymmetric background potential
of Allen \& Santill\'an (\cite{AS91}) has been scaled to give a
rotation velocity of 254 km$\,$s$^{-1}$ at the solar position, based
on the most-recent radio astrometry observations by Reid et al.\
({\cite{RMZ09}).  We keep the original value $R_0$=8.5 kpc of the
solar galactocentric distance (Reid et al.\ ({\cite{RMZ09}) give
$R_0$=8.4$\pm$0.6 kpc). The adopted parameters for the bar and spiral
arms and the corresponding references are provided in Table
\ref{tbl-1}.  The values of the parameters are based on recent
observations of the Milky Way. The bar model is an inhomogeneous
ellipsoidal potential that closely approximates model S of
Freudenreich (\cite{F98}) from the COBE/DIRBE data of the Galactic
bar. For the spiral perturbation, Pichardo et al.\ (\cite{PMME03})
refined their model until self-consistent orbital solutions were
found. The orbital self-consistency of the spiral arms was tested
through the reinforcement of the spiral potential by the stellar
orbits (Patsis et al.\ \cite{PCG91}).  For an extensive description of
the models, see Pichardo et al.\ (\cite{PMME03}, \cite{PMM04}).

\begin{table*}[t!]
\begin{center}
\caption{{Local standard of rest initial velocity $(U,V,W)$ and Galactic ($\Pi,
\Theta$) (km$\,$s$^{-1}$) for the three heliocentric distances $d$
(kpc).}}
\label{tbl-2}
\begin{tabular}{cccccc}
\hline\hline
&&&&&\\
{$d$} & {$U$} & {$V$} & {$W$} &
{$\Pi$} & {$\Theta$}\\
&&&&&\\
\hline
&&&&&\\
0.815 & $31.92\pm3.4$ & $-21.66\pm3.7$ & $-8.71 \pm4.3$ & $21.55\pm3.2$  & $233.53\pm3.8$ \\
0.765 & $30.77\pm3.0$ & $-21.03\pm3.4$ & $-6.64 \pm3.6$ & $20.98\pm3.0$  & $234.05\pm3.5$ \\
0.865 & $33.06\pm3.4$ & $-22.28\pm3.9$ & $-10.78\pm4.1$ & $22.12\pm3.4$  & $233.02\pm4.0$ \\
&&&&&\\
\hline
\end{tabular}
\end{center}
\end{table*}

\begin{table*}[t!]
\begin{center}
\caption{{Orbit parameters 
in the scaled axisymmetric potential
for the three heliocentric distances.}} 
\label{tbl-3}
\begin{tabular}{cccccccccc}
\hline\hline &&&&&&&&&\\ $d$ &
$r_{\rm min}$ & $r_{\rm max}$ & $z_{\rm max}$ & $e$ & $P_\phi$$$ & $P_r$ & $P_z$ & $h$ & $E$\\ 
(kpc) & (kpc) & (kpc) & (kpc) & & (Myr) & (Myr) & (Myr) & (kpc$\,$km$\,$s$^{-1}$) & (10$^2$ km$^2$$\,$s$^{-2}$)\\ 
&&&&&&&&&\\ 
\hline 
&&&&&&&&&\\
0.815 & 7.65& 9.28 & 0.46& 0.096 & 207& 146 & 76 & 2118.3& $-$1589.38 \\
0.765 & 7.65& 9.24 & 0.42& 0.094 & 206& 145 & 74 & 2114.7& $-$1591.50 \\
0.865 & 7.66& 9.33 & 0.50& 0.098 & 208& 147 & 78 & 2122.0& $-$1587.21 \\
&&&&&&&&&\\ \hline
\end{tabular}
\end{center}
\end{table*}

\begin{figure*}[t!]
\centering
\includegraphics[width=8.45cm]{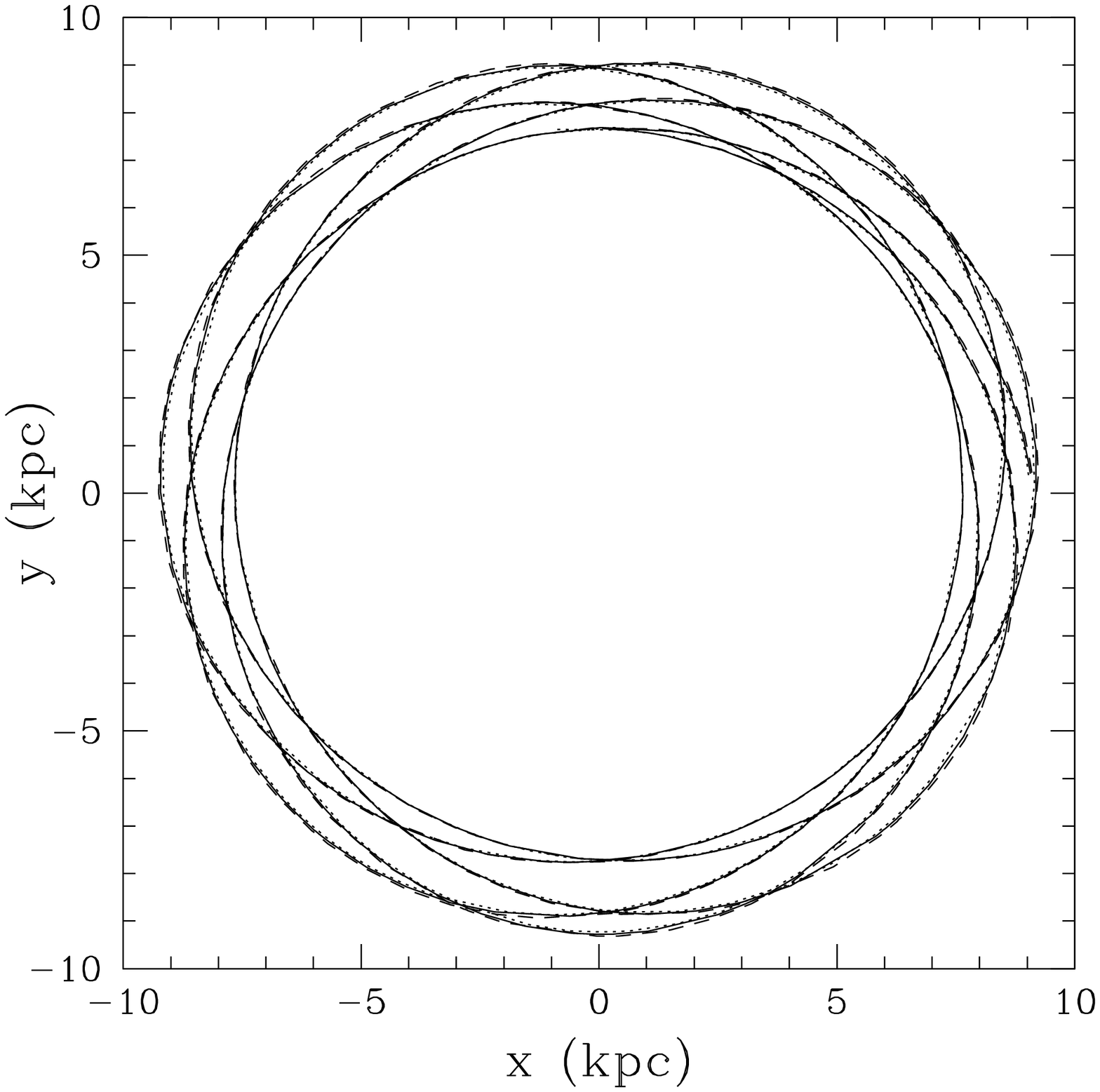}
\includegraphics[width=8.45cm]{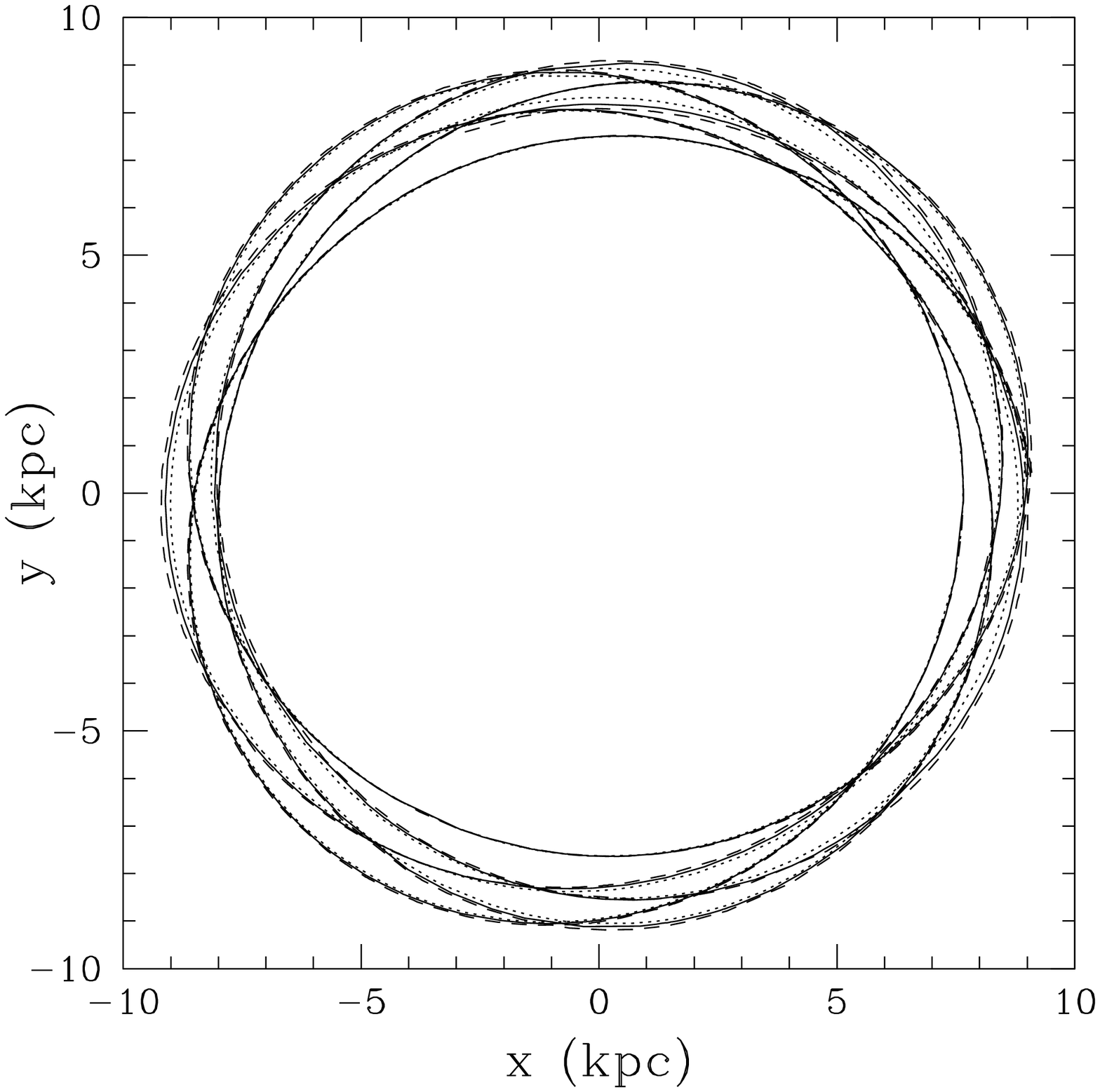}
\caption{Projection of the orbit on the Galactic plane computed in the
scaled axisymmetric potential (\textit{left}), and in the bar + spiral
arms scaled Galactic potential (\textit{right}).  We adopted an
heliocentric distance $d$=0.815 kpc (full line), 0.765 kpc (dotted
line), and 0.865 kpc (dashed line).}
\label{fig23}
\end{figure*}

Table~\ref{tbl-2} gives the local standard of rest (LSR) initial
velocity $(U,V,W)$ and the corresponding cylindrical components
$(\Pi,\Theta)$ for three different heliocentric distances of M 67
(central, minimum, and maximum).  We used the solar motion
$(U,V,W)$=$(-10, 5.2, 7.2)$ km$\,$s$^{-1}$ (Dehnen \& Binney
\cite{DB98}). $U$ is negative toward the Galactic center.  In Table
\ref{tbl-3} we give some parameters of the Galactic orbits,
corresponding to the three heliocentric distances, computed backwards
in time during 1 Gyr in the scaled axisymmetric potential.  Columns 2
and 3 show the minimum and maximum galactocentric distances and
Col.\ 4 the maximum $z$-distance from the Galactic plane; the orbital
eccentricity is given in Col.\ 5; Cols.\ 6 to 8 give the azimuthal,
radial, and vertical periods respectively, and the last two columns
the $z$-component of the angular momentum and energy per unit mass.

Previous computations of the Galactic orbit of M~67 were made by
Keenan et al.\ (\cite{KIH73}) and Allen \& Martos (\cite{AM88}) in
other axisymmetric potentials. Keenan et al. (\cite{KIH73}) used the
Galactic models of Schmidt (\cite{S56}) (with an $R_0$ value of 8.2
kpc and a circular velocity at the Sun's position $V_0$=216 km\
s$^{-1}$) and of Innanen (\cite{I66}) (with an $R_0$=10 kpc and a
$V_0$=250 km s$^{-1}$).  They obtained $R_{\rm max}$$=$8.8 kpc and
10.7 kpc respectively $z_{\rm max}$$=$ 0.4 kpc and an eccentricity of
0.1 for both models.  Allen \& Martos (\cite{AM88}) used their
galactic model (Allen \& Martos \cite{AM86}) with $R_0$=8 kpc and
$V_0$=225 km s$^{-1}$.  They obtained $R_{\rm max}$=8.7 kpc, $z_{\rm
max}$=0.48 kpc and an eccentricity of 0.11.  In all cases, the orbit
of M~67 has a small eccentricity, and the differences found are mostly
attributable to the different Galactic parameters and values for the
solar motion employed.  Indeed, the different values obtained for
$R_{\rm max}$ result mostly from the scaling of the Sun-center
distance, as shown by the ratios $R_0/R_{\rm max}$, which are all
between 1.07 and 1.09.  The more contemporary, much-improved Galactic
models and the precise value now available for the absolute proper
motion of M~67 should result in a more reliable orbit for this
cluster.

We also computed the Galactic orbit of M~67 in the non-axisymmetric
potential, first including only the bar, then with the bar + spiral
arms.  The scaled background axisymmetric potential was considered.
The angular velocity of the bar was taken as ${\Omega}_B$=60
km$\,$s$^{-1}$$\,$kpc$^{-1}$; the other parameters of the
non-axisymmetric components are given in Table~\ref{tbl-1}. In the
case with spiral arms, the mass of these arms was taken as 2.2\% the
mass of the scaled disk component, which is 3\% of the mass of the
original disk.  This mass gives a force contrast as listed in Table
\ref{tbl-1}.

Figure~\ref{fig1} shows the meridional orbits computed in the
axisymmetric potential (left panels), axisymmetric + bar potential
(central panels) and axisymmetric + bar + spiral arms potential (right
panels), using the three heliocentric distances. As shown, the
potential that includes only the Galactic bar gives an orbit similar
to that obtained with the axisymmetric potential.  This is because the
orbit of M~67 lies far outside the region of the bar. However, the
potential that includes both the bar and spiral arms shows a different
behavior. The orbit is perturbed by the spiral arms, mainly in the
radial direction.  The radial dispersion is not very strong in M~67,
but distorts what would be a box orbit. Thus, a moderate spiral
potential has important effects in the kinematics of orbits near the
Galactic plane, as is the case in M~67, and in general, as is the case
for the solar neighborhood stars (Antoja et al.\ \cite{AV09}). This
result holds for the allowed variations in proper motion, radial
velocity and distance, because the most important parameter that
affects the orbit is the mass of the spiral arms.

In left panel of Fig.~\ref{fig23} we show the projection of the orbit
on the Galactic plane, computed in the axisymmetric potential. At the
scale shown in this figure there is no appreciable difference between
the orbits using the three heliocentric distances. The right panel
shows the corresponding orbits in the axisymmetric + bar + spiral arms
potential. There is a slight difference in the azimuthal behavior in
both figures.

We have also computed the tidal radius of M~67 in the axisymmetric +
bar + spiral arms scaled potential. With a total mass in M~67 of 279
$M_{\odot}$, listed by Piskunov et al.\ (\cite{PSK08}), using King's
equation (King \cite{K62}) we obtained a mean tidal radius of 7.1
pc. With the alternative equation (1) in Allen et al.\ (\cite{AMP06})
the tidal radius is 8 pc.  Both results are near the 9.6 pc value
listed by Piskunov et al.\ (\cite{PSK08}).

\begin{acknowledgements}
We thank the anonymous referee for a careful reading of the
manuscript and useful comments.  A.B.\ acknowledges the support by
the CA.RI.PA.RO.  foundation, and by the STScI under the
\textit{``2008 graduate research assistantship''} program.
B.P. thanks project UNAM through grant PAPIIT IN1 19708. 
G.P.\ acknowledges partial support by PRIN07 (prot.\ 20075TP5K9)
and by ASI under the program ASI-INAF I/016/07/0 program.
\end{acknowledgements} 

%%%%%%%%%%%%%%%%%%%%%%%%%%%%%%%%%%%%%%%%%%%%%%%%%%%%%%%%%%%%%%%%%%%%%%

\end{document}